\documentclass[11pt,a4paper]{article}
\usepackage{jheppub}

\usepackage{amsfonts}
\usepackage{amssymb}
\usepackage{amsmath}
\usepackage{color}
\usepackage[small,bf]{caption}
\usepackage[usenames,dvipsnames]{xcolor}
\usepackage{subfigure}
\usepackage{verbatim}
\usepackage{dsfont}
\usepackage{tikz}
\usepackage{braket}
\usepackage{lscape}
\usepackage{slashed}
\usepackage{booktabs}
\usetikzlibrary{shapes,arrows}
\usepackage{mathtools}
\usepackage{graphicx}
\usepackage{fontawesome}
\usepackage{enumitem}
\usepackage{multirow}

\def\var{\mbox{\boldmath $\varepsilon$}}
\def\r{\mbox{{\bf  r}}}
\def\p{\mbox{\boldmath $p$}}
\def\q{\mbox{\boldmath $q$}}
\def\k{\mbox{\boldmath $k$}}
\def\t{\mbox{\boldmath $t$}}

\allowdisplaybreaks[1]

\numberwithin{equation}{section}

\begin{document}

\title{Analysis of flux-integrated semi-exclusive cross sections for charged
  current quasi-elastic neutrino scattering off  $\boldmath{{}^{40}}$Ar at
  energies available at the MicroBooNE experiment}
\author{A.~V.~Butkevich}
\affiliation{Institute for Nuclear Research,
Russian Academy of Sciences, Moscow 117312, Russia\\}
\emailAdd{butkevic@inr.ru}

\date{\today}

\abstract{
Flux-integrated semi-exclusive differential and integral cross sections for
quasi-elastic neutrino charged-current scattering on argon are analyzed. We
calculate these cross sections using the relativistic distorted-wave impulse
approximation and compare with recent MicroBooNE data. We found that the
measured cross sections can be described well within the experimental
uncertainties with value of the nucleon axial mass $1 < M_A < 1.2$ GeV.
The contribution of the exclusive channel $\boldmath{(\nu_{\mu}, \mu p)}$ to
the flux-integrated inclusive cross sections is about 50\%.}



\maketitle
\section{Introduction}
Current~\cite{Acero:2019ksn, Abe:2020vdv} and
future~\cite{Acciarri:2016crz, Abe:2018uyc} neutrino
oscillation experiments use high-intensity muon-(anti)neutrino beams that are
not monoenergetic and peak in the energy range from tens of MeV to a few
GeV. The goal of these experiments is to measure oscillation features in
the neutrino energy spectrum reconstructed at far detectors. To evaluate the
oscillation parameters, the probabilities of neutrino oscillations as
functions of neutrino energy are measured. The accuracy to which they can
extract neutrino oscillation parameters depends on their ability to determine
the individual energy of detected neutrino. This requires detailed
understanding of neutrino interactions with nuclei. 

In the energy range $\varepsilon_{\nu} \sim 0.2$ -- 5 GeV  charged-current
(CC) quasielastic (QE) scattering and scattering induced by two-body meson
exchange current (MEC), resonance production and deep inelastic process yield
the main contributions to the neutrino-nucleus interaction. The understanding
of these interactions comes through cross sections measurements on various
channels. The incident neutrino energy can be reconstructed using the
calorimetric method, which rely not only on the lepton and hadron energies
visible in the final state after the neutrino has interacted but also on
models of the neutrino-nucleus interactions that are implement in neutrino
events generators.

The CCQE interaction forms a significant contribution in many accelerator-based
neutrino experiments~\cite{Aguilar-Arevalo:2010zc, Abe:2014iza,
  Fiorentini:2013ezn, Walton:2014esl, Betancourt:2017uso, Abe:2018pwo,
  Abratenko:2019jqo, Abratenko:2020acr, Abratenko:2020sga}. Because the CCQE
interaction represents a two
particle scattering process, its final state topology is simple with an
easy identifiable lepton, and neutrino energy may be estimated using the
outgoing lepton kinematics, i.e. applying kinematical methods. However, as
neutrino beams have broad energy distributions, various contributions to the
inclusive cross section, where only final lepton is detected, can significantly
overlap with each other making it difficult to identify the channels of neutrino
interaction. The interpretation of the inclusive CCQE data is complicated
because of the presence of other interactions such as MEC and pion production,
where the pion is absorbed in the residual nucleus. So, when only the muon is
detected, the event can easily be mistaken as a CCQE interaction and 
application of the kinematic method will lead to a bias in the neutrino energy
estimation.

Compared to inclusive experiments, semi-exclusive scattering provides 
additional information about hadrons in the final state. In this process
the neutrino removes a single intact nucleon from the nucleus without
production of any
additional particles. For these CC1p0$\pi$ events the experimental signature
requires the identification of a neutrino interaction vertex with an
outgoing lepton, exactly one outgoing proton, and no additional
particles, which is
relatively straight forward to measure. The information about hadrons will
improve the accuracy of reconstruction of the incoming neutrino energy.
Understanding the interaction of neutrino with argon nuclei is of particular
importance, since neutrino oscillation experiments such as
DUNE~\cite{Acciarri:2016crz} and SBN~\cite{Antonello:2015lea} employ
neutrino detectors using liquid
argon time projector chamber (LArTPs).

Weak interactions of neutrinos probe the nucleus in a similar way as
electromagnetic electron interactions. Of course, there are a number of
differences with neutrino scattering, the most important one being the absence
of the axial current contribution. Nevertheless the influence of the nuclear medium
is the same as in neutrino-nucleus scattering data. Precise electron-scattering
data  gives a unique opportunity to validate the nuclear model employed in
neutrino physics. Therefore, the detailed comparison with $(e,e'p)$ data is
a necessary test for any theoretical model used to describe the CC1p0$\pi$ cross
sections.

Systematic measurements of $(e,e'p)$ cross sections were performed at
Saclay~\cite{Frullani:1984nn}, NIKHEF~\cite{Dieperink:1990uk}, and
JLab~\cite{Fissum:2004we, Dutta:2003yt, Gu:2020shc}.
The impact of the MEC on the nuclear response functions and differential cross
section for the knockout of protons from ${}^{16}$O was studied in
Ref.~\cite{Fissum:2004we}. It was shown that explicit inclusion of the two-body
current contribution does not markedly improve the overall agreement between
the data and calculated cross sections. Unfortunately, the cross section data
for the semi-exclusive lepton scattering on argon in the relevant energy range are
rather scare. There are only experimental data for $\approx 2.2$ GeV electron
scattering of ${}^{40}$Ar~\cite{Gu:2020shc} and flux-integrated differential  
CC1p0$\pi$ cross sections for $\nu_{\mu}{}^{40}$Ar scattering, measured with
the MicroBooNE detector~\cite{Abratenko:2020sga}. On the other hand the structure of
${}^{40}$Ca and ${}^{40}$Ar nuclei are almost identical, and for calcium high
resolution exclusive $(e,e'p)$ experiments were carried out at
Tokyo~\cite{Nakamura:1974zz, Nakamura:1976kn}, Saclay~\cite{Mougey:1976sc}, and
NIKHEF~\cite{Kramer:1990,Kramer:1989uiu,Kramer:2000kc}.

The data analysis of $(e,e'p)$ and $(\nu_{\mu},\mu p)$ processes was performed
in Refs.~\cite{Fissum:2004we, Kelly:1996hd, Udias:2001tc, Kelly:2005is, Maieron:2003df,
  Meucci:2003cv, Martinez:2005xe} within the
relativistic distorted-wave impulse approximation (RDWIA)~\cite
{Picklesimer:1985ey, Picklesimer:1986wj, Kelly:1998ti}, using a relativistic
shell model approach. The implementation of the
final state interaction (FSI) of the ejected nucleon has been done differently.
The RDWIA approach describes with high degree of accuracy the experimental shape
of the outgoing particle momentum distributions. In order to reproduce
experimental cross sections, normalizations of the bound-state wave functions
were fit to the data and identified with the spectroscopic factors.
The semi-inclusive process for CCQE reactions was recently discussed in Refs.~
\cite{Gonzalez-Jimenez:2021ohu, Amaro:2021sec}.

In this work we calculated the flux-integrated differential cross sections of
${}^{40}$Ar$(\nu_{\mu}, \mu p)$ interactions with the RDWIA approach, using the
Booster Neutrino Beam (BNB) at Fermilab~\cite{Aguilar-Arevalo:2008yp}. This
approach was successfully applied in Refs.~\cite{Butkevich:2007gm,
  Butkevich:2009cp, Butkevich:2010cr, Butkevich:2012zr, Butkevich:2020ysf} for
calculation of the CCQE semi-exclusive and inclusive cross sections for the
electron and neutrino scattering on ${}^{12}$C, ${}^{16}$O, ${}^{40}$Ca, and
${}^{40}$Ar nuclei. The aim of this work is to test the RDWIA predictions
against the MicroBooNE data~\cite{Abratenko:2020sga}. Within this model we
estimated the range of the values of nucleon axial mass from the measured
CC1p0$\pi$ flux-integrated differential cross sections.

The outline of this article is as follows. In Sec.II we present briefly the
formalism for the CCQE semi-inclusive scattering process and basic aspects of
the RDWIA approach, used for the calculation. The results are presented and
discussed in Sec.III. Our conclusions are summarized in Sec.IV.

\section{Formalism of quasi-elastic scattering and RDWIA}

In this section we consider shortly the formalism used to describe 
electron and neutrino quasi-elastic exclusive
\begin{equation}\label{qe:excl}
l(k_i) + A(p_A)  \rightarrow l^{\prime}(k_f) + N(p_x) + B(p_B),      
\end{equation}
and inclusive
\begin{equation}\label{qe:incl}
l(k_i) + A(p_A)  \rightarrow l^{\prime}(k_f) + X                      
\end{equation}
scattering off nuclei in the one-photon (W-boson) exchange approximation. 
Here $l$ labels the incident lepton [electron or muon (anti)neutrino], and
$l^{\prime}$ represents the scattered lepton (electron or muon),
$k_i=(\varepsilon_i,\k_i)$ 
and $k_f=(\varepsilon_f,\k_f)$ are the initial and final lepton 
momenta, $p_A=(\varepsilon_A,\p_A)$, and $p_B=(\varepsilon_B,\p_B)$ are 
the initial and final target momenta, $p_x=(\varepsilon_x,\p_x)$ is the 
ejectile nucleon momentum, $q=(\omega,\q)$ is the momentum transfer carried by 
the virtual photon (W-boson), and $Q^2=-q^2=\q^2-\omega^2$ is the photon 
(W-boson) virtuality.

\subsection{CCQE lepton-nucleus cross sections}

In the laboratory frame the differential cross section for exclusive
electron ($\sigma ^{el}$) and (anti)neutrino ($\sigma ^{cc}$) CC scattering can
be written as
\begin{subequations}
\begin{align}
\frac{d^6\sigma^{el}}{d\varepsilon_f d\Omega_f d\varepsilon_x d\Omega_x} &=
\frac{\vert\p_x\vert\varepsilon_x}{(2\pi)^3}\frac{\varepsilon_f}{\varepsilon_i}
 \frac{\alpha^2}{Q^4} L_{\mu \nu}^{(el)}\mathcal{W}^{\mu \nu (el)}
\\                                                                  
\frac{d^6\sigma^{cc}}{d\varepsilon_f d\Omega_f d\varepsilon_x d\Omega_x} &=
\frac{\vert\p_x\vert\varepsilon_x}{(2\pi)^5}\frac{\vert\k_f\vert}
{\varepsilon_i} \frac{G^2\cos^2\theta_c}{2} L_{\mu \nu}^{(cc)}
\mathcal{W}^{\mu \nu (cc)},
\end{align}
\end{subequations}
 where $\Omega_f$ is the solid angle for the lepton momentum, $\Omega_x$ is the
 solid angle for the ejectile nucleon momentum, $\alpha\simeq 1/137$ is the
fine-structure constant, $G \simeq$ 1.16639 $\times 10^{-11}$ MeV$^{-2}$ is
the Fermi constant, $\theta_C$ is the Cabbibo angle
($\cos \theta_C \approx$ 0.9749), $L^{\mu \nu}$ is the lepton tensor, and
 $\mathcal{W}^{(el)}_{\mu \nu}$ and $\mathcal{W}^{(cc)}_{\mu \nu}$ are
correspondingly the electromagnetic and weak CC nuclear tensors.

For exclusive reactions in which only a single discrete state or narrow
resonance of the target is excited, it is possible to integrate over the
peak in missing energy and obtain a fivefold differential cross section of
the form
\begin{subequations}
\begin{align}
\label{cs5:el}
\frac{d^5\sigma^{el}}{d\varepsilon_f d\Omega_f d\Omega_x} &= R
\frac{\vert\p_x\vert\tilde{\varepsilon}_x}{(2\pi)^3}\frac{\varepsilon_f}
{\varepsilon_i} \frac{\alpha^2}{Q^4} L_{\mu \nu}^{(el)}W^{\mu \nu (el)}
\\                                                                       
\label{cs5:cc}
\frac{d^5\sigma^{cc}}{d\varepsilon_f d\Omega_f d\Omega_x} &= R
\frac{\vert\p_x\vert\tilde{\varepsilon}_x}{(2\pi)^5}\frac{\vert\k_f\vert}
{\varepsilon_i} \frac{G^2\cos^2\theta_c}{2} L_{\mu \nu}^{(cc)}W^{\mu \nu (cc)},
\end{align}
\end{subequations}
where $R$ is a recoil factor
  \begin{equation}\label{Rec}
R =\int d\varepsilon_x \delta(\varepsilon_x + \varepsilon_B - \omega -m_A)=
{\bigg\vert 1- \frac{\tilde{\varepsilon}_x}{\varepsilon_B}
\frac{\p_x\cdot \p_B}{\p_x\cdot \p_x}\bigg\vert}^{-1},                    
\end{equation}
$\tilde{\varepsilon}_x$ is solution to equation
$
\varepsilon_x+\varepsilon_B-m_A-\omega=0,
$
where $\varepsilon_B=\sqrt{m^2_B+\p^2_B}$, $~\p_B=\q-\p_x$ and $m_A$ and $m_B$
are masses of the target and recoil nucleus, respectively. Note, that missing
momentum is $\p_m=\p_x-\q$ and missing energy $\var_m$ is defined by
$\var_m=m + m_B -m_A$. The differential cross sections
($d^3\sigma^{el(cc)}/d\var_fd\Omega_f)_{ex}$ 
can be obtained by integrating the exclusive cross sections (\ref{cs5:el}) and
(\ref{cs5:cc}) over solid angle for the ejectile nucleon.

All information about the nuclear structure and FSI effects is 
contained in the electromagnetic and weak CC hadronic tensors,
$W^{(el)}_{\mu \nu}$ and $W^{(cc)}_{\mu \nu}$, which are given by the bilinear 
products of the transition matrix elements of the nuclear electromagnetic or CC 
operator $J^{(el)(cc)}_{\mu}$ between the initial nucleus state 
$\vert A \rangle $ and the final state $\vert B_f \rangle$ as
\begin{eqnarray}
W^{(el)(cc)}_{\mu \nu } &=& \sum_f \langle B_f,p_x\vert                     
J^{(el)(cc)}_{\mu}\vert A\rangle \langle A\vert
J^{(el)(cc)\dagger}_{\nu}\vert B_f,p_x\rangle,              
\label{W}
\end{eqnarray}
where the sum is taken over undetected states.

In the inclusive reaction (\ref{qe:incl}) only the outgoing lepton is detected and
lepton scattering cross sections in term of nuclear response functions can be
written as 
\begin{subequations}
\begin{align}
\frac{d^3\sigma^{el}}{d\varepsilon_f d\Omega_f} &=
\sigma_M\big(V_LR^{(el)}_L + V_TR^{(el)}_T\big),
\\                                                                       
\frac{d^3\sigma^{cc}}{d\varepsilon_f d\Omega_f} &=
\frac{G^2\cos^2\theta_c}{(2\pi)^2} \varepsilon_f
\vert \k_f \vert\big ( v_0R_0 + v_TR_T
+ v_{zz}R_{zz} -v_{0z}R_{0z}- hv_{xy}R_{xy}\big),
\end{align}
\end{subequations}
where
\begin{equation}
\sigma_M = \frac{\alpha^2\cos^2 \theta/2}{4\varepsilon^2_i\sin^4 \theta/2} 
\end{equation}
is the Mott cross section. The coupling coefficient $V_k$ and $v_k$, the
expression of which are given in Ref.~\cite{Butkevich:2007gm} are kinematic
factors depending on the lepton's kinematics. The response functions $R_i$
are given in terms of components of the inclusive hadronic tensors~
(\cite{Butkevich:2007gm}) and depend on the variables
$(Q^2, \omega)$ or $(|\q|,\omega)$.

The experimental data of the $(e,e'p)$ reaction are usually presented in terms 
of the reduced cross section
\begin{equation}
\sigma_{red} = \frac{d^5\sigma}{d\varepsilon_f d\Omega_f d\Omega_x}
/K^{(el)(cc)}\sigma_{lN},                                                 
\end{equation}
where
$K^{el} = R {p_x\varepsilon_x}/{(2\pi)^3}$ and
$K^{cc}=R {p_x\varepsilon_x}/{(2\pi)^5}$
are phase-space factors for electron and neutrino scattering  and 
$\sigma_{lN}$ is the corresponding
elementary cross section for the lepton scattering from the moving free
nucleon. The reduced cross section is an interesting quantity that can be 
regarded as the nucleon momentum distribution modified by FSI. Therefore   
these cross sections for (anti)neutrino  scattering off nuclei are similar
to the electron scattering apart from small differences at low beam energy
due to effects of Coulomb distortion of the incoming electron wave function.
Precise electron reduced cross section data can be used to validate the
 neutrino reduced cross sections employed in neutrino generators.

\subsection{Model}

We describe genuine QE electron-nuclear scattering within the RDWIA
approach. This formalism is based on the impulse approximation (IA), assuming
that the incoming lepton interacts with only one nucleon of the target, which
is subsequently emitted.
In this approximation the nuclear current is written as a sum of
single-nucleon currents and the nuclear matrix element in Eq.~(\ref{W}) takes the
form 
\begin{eqnarray}\label{Eq12}
\langle p,B\vert J^{\mu}\vert A\rangle &=& \int d^3r~ \exp(i\t\cdot\r)
\overline{\Psi}^{(-)}(\p,\r)
\Gamma^{\mu}\Phi(\r),                                                     
\end{eqnarray}
where $\Gamma^{\mu}$ is the vertex function, $\t=\varepsilon_B\q/W$ is the
recoil-corrected momentum transfer, $W=\sqrt{(m_A + \omega)^2 - \q^2}$ is the
invariant mass and $\Phi$ and $\Psi^{(-)}$ are relativistic bound-state and
outgoing wave functions.

For electron scattering, we use the CC2 electromagnetic vertex
function for a free nucleon~\cite{DeForest:1983ahx}
\begin{equation}
\Gamma^{\mu}_V = F_V(Q^2)\gamma^{\mu} + {i}\sigma^{\mu \nu}\frac{q_{\nu}}
{2m}F_M(Q^2),                                                          
\end{equation}
where $\sigma^{\mu \nu}=i[\gamma^{\mu},\gamma^{\nu}]/2$, $F_V$ and
$F_M$ are the Dirac and Pauli nucleon form factors.
The single-nucleon charged current has $V{-}A$ structure $J^{\mu(cc)} = 
J^{\mu}_V + J^{\mu}_A$.
For a free-nucleon vertex function 
$\Gamma^{\mu(cc)} = \Gamma^{\mu}_V + \Gamma^{\mu}_A$ we use the CC2 vector 
current vertex function
\begin{equation}
\Gamma^{\mu}_V = F_V(Q^2)\gamma^{\mu} + {i}\sigma^{\mu \nu}\frac{q_{\nu}}
{2m}F_M(Q^2)                                                          
\end{equation}
and the axial current vertex function
\begin{equation}
\Gamma^{\mu}_A = F_A(Q^2)\gamma^{\mu}\gamma_5 + F_P(Q^2)q^{\mu}\gamma_5. 
\end{equation}
The weak vector form factors $F_V$ and $F_M$ are related to the corresponding 
electromagnetic form factors $F^{(el)}_V$ and $F^{(el)}_M$ for protons and 
neutrons by the hypothesis of the conserved vector current. We use the 
approximation of Ref.~\cite{Mergell:1995bf} for the Dirac and Pauli nucleon
form factors. Because the bound nucleons are off-shell we employ the de Forest 
prescription~\cite{DeForest:1983ahx} and Coulomb gauge for the off-shell
vector current vertex $\Gamma^{\mu}_V$. The vector-axial $F_A$ and pseudoscalar
$F_P$ form factors are parametrized using a dipole approximation: 
\begin{equation}
F_A(Q^2)=\frac{F_A(0)}{(1+Q^2/M_A^2)^2},\quad                         
F_P(Q^2)=\frac{2m F_A(Q^2)}{m_{\pi}^2+Q^2},
\end{equation}
where $F_A(0)=1.2724$, $M_A$ is the axial mass that controls $Q^2$-dependence 
of $F_A(Q^2)$, and $m_\pi$ is the pion mass.

In the RDWIA calculations the independent particle shell model (IPSM) is
assumed for the nuclear structure. In Eq.(\ref{Eq12}) the relativistic
bound-state wave functions for nucleons $\Phi$ are obtained as the 
self-consistent solutions of a Dirac equation, derived 
within a relativistic mean-field approach from a Lagrangian containing 
$\sigma$, $\omega$, and $\rho$ mesons~\cite{Horowitz:1981xw}. These functions
were calculated by the TIMORA code~\cite{Horowitz:1991}
with the normalization factors $S_{\alpha}$ relative to full occupancy of the
IPSM orbital $\alpha$ of ${}^{40}$Ca. For ${}^{40}$Ca and ${}^{40}$Ar an average
factor $\langle S \rangle \approx 87\%$. This estimation of depletion of hole
states follows from the RDWIA analysis of ${}^{40}$Ca$(e,e'p)$
data~\cite{Butkevich:2012zr}. The source of the reduction of the
$(e,e'p)$ spectroscopic factors with respect to the mean field values are 
the short-range and tensor correlations in the ground state, leading to the
appearance of the high-momentum and high-energy component in the nucleon
distribution in the target. Mean values of the proton and neutron binding
energies and occupancies of shells are given also in
Ref.~\cite{Butkevich:2012zr}. 
\begin{figure*}
  \begin{center}
    \includegraphics[height=10cm,width=11cm]{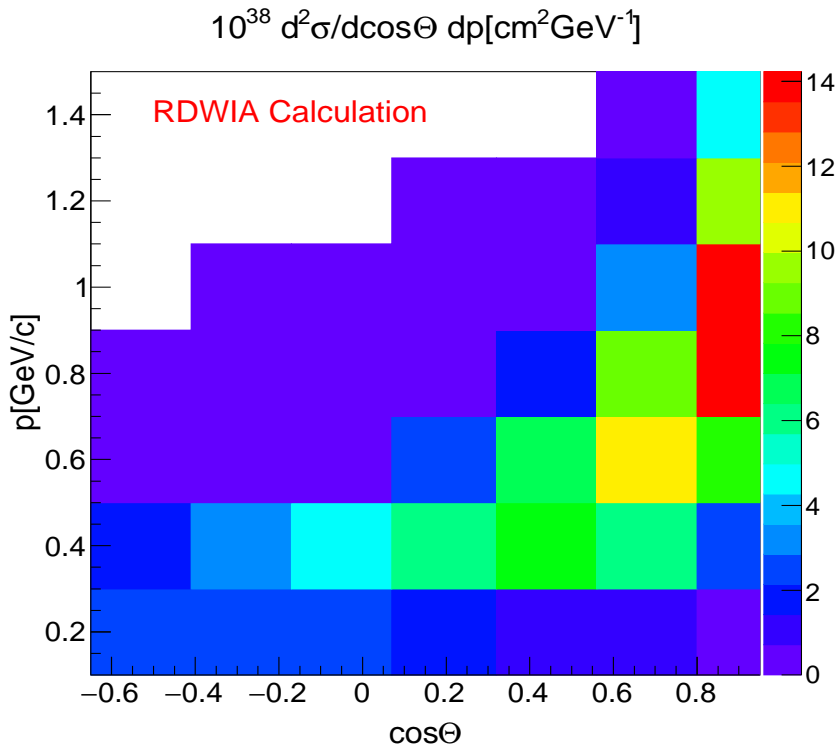}
  \end{center}
  \caption{\label{Fig1} The flux-integrated double-differential CC1p0${\pi}$
    cross section as a function of muon momentum and the cosine of the muon
    scattering angle.
}
\end{figure*}

In the RDWIA model, final state interaction effects for the outgoing nucleon
are taken into account.  The system of two coupled first-order Dirac equations
is reduced to a single second-order Schr\"odinger-like equation for the upper
component of the Dirac wave function $\Psi$. This equation contains equivalent
nonrelativistic central and spin-orbit potentials which are functions of the
relativistic, energy dependent, scalar, and vector optical potentials. 
The optical potential consists of areal part, which describes the rescattering
of the ejected nucleon and an imaginary part which account for its absorption
into unobserved channels. We use the LEA program~\cite{Kelly:1996}
for the numerical calculation of the distorted wave functions with the EDAD1
parametrization~\cite{Cooper:1993nx} of the relativistic optical potential for
calcium. This code was successfully tested in Ref.~\cite{Butkevich:2012zr}
against $A(e,e'p)$ data for electron scattering off ${}^{40}$Ca.
In Ref.~\cite{Butkevich:2012zr} the reduced cross
sections as functions of missing momentum calculated in the RDWIA approach
for the ${}^{40}$Ca$(e,e'p)$ reaction are shown in figures 1 and 2 with NIKHEF
data and provide a good description of the measured distributions. Neutrino
and antineutrino cross sections are also shown in figure 1 for comparison.
\begin{figure*}
  \begin{center}
    \includegraphics[height=16cm,width=16cm]{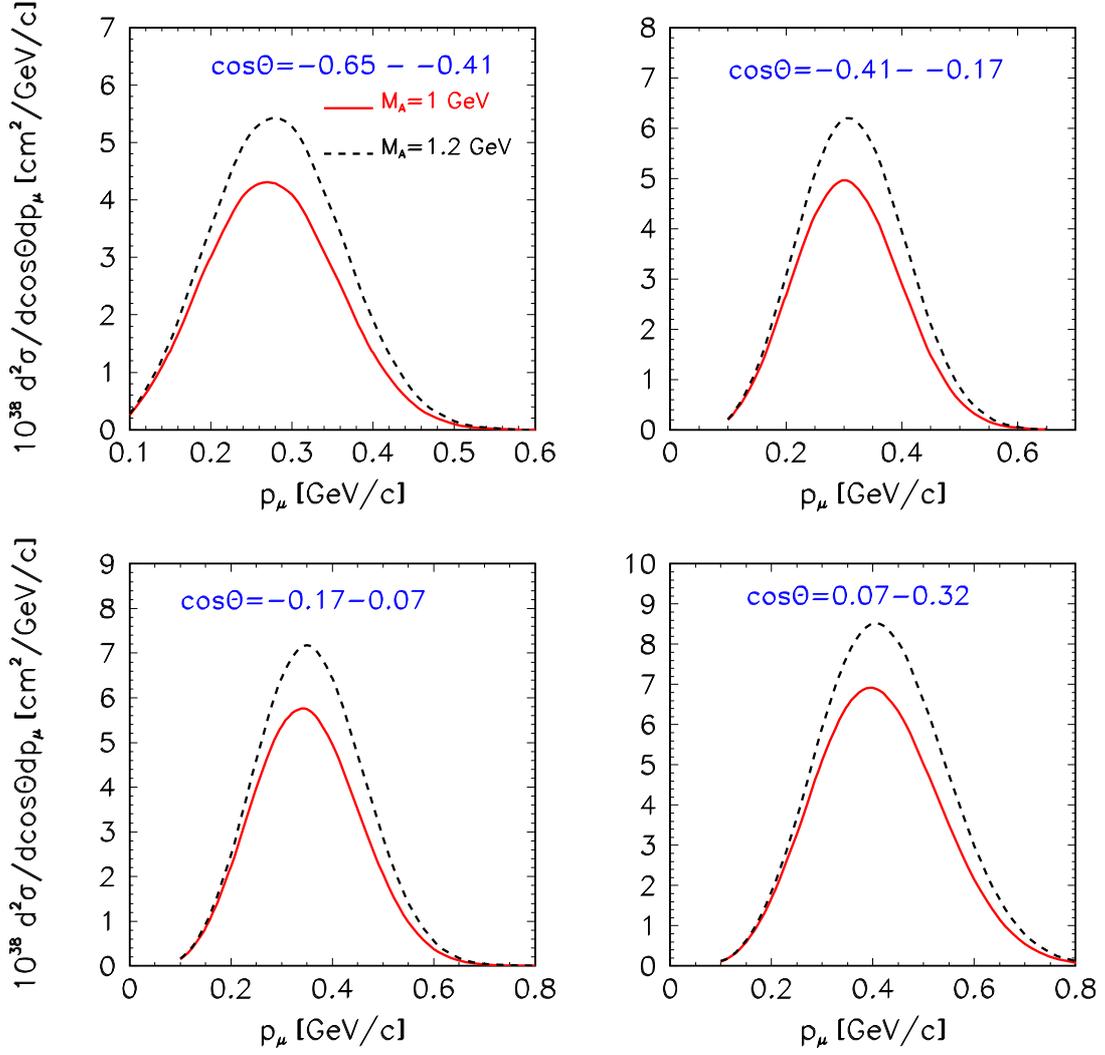}
  \end{center}
  \caption{\label{Fig2} The flux-integrated semi-exclusive CCQE 
    $d^2\sigma/dp_{\mu}d\cos\theta $ cross section for $\nu_{\mu}-{}^{40}Ar$
    scattering as a function of $p_{\mu}$ for the four muon scattering angle
    bins: $\cos\theta=[(-0.65) - (-0.4)], [(-0.4) - (-0.17)],
    [(-0.17) - 0.07]$ and $(0.07 - 0.32)$. As shown in the key, cross
      sections were calculated with $M_A=1$ GeV and 1.2 GeV. }
\end{figure*}
 
The inclusive cross sections with the FSI effects, taking into account
short-range nucleon-nucleon ($NN$) correlations were calculated using the
method proposed in Ref.~\cite{Butkevich:2007gm} with the nucleon high-momentum
and high-energy distribution from Ref.~\cite{CiofidegliAtti:1995qe} that was
renormalized to value of 13\% for calcium and argon. The contribution of the
$NN$-correlated pairs is evaluated in impulse approximation, i.e., the virtual
photon couples to only one member of the $NN$-pair.
\begin{figure*}
  \begin{center}
    \includegraphics[height=9cm,width=18cm]{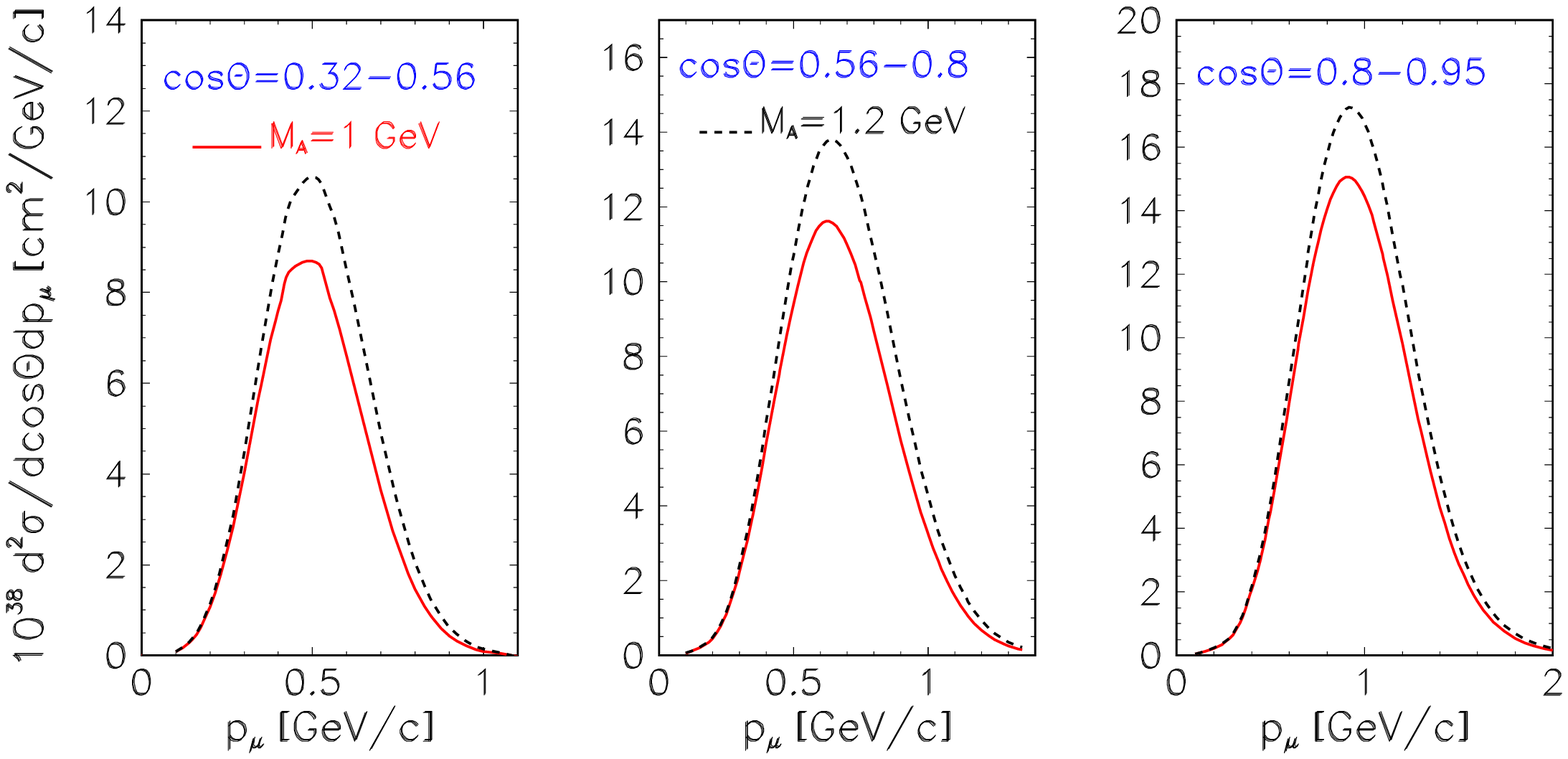}
  \end{center}
  \caption{\label{Fig3} Same as figure 2 but for muon scattering angle bins:
    $\cos\theta=(0.32 - 0.56), (0.56 - 0.8), (0.8 - 0.95)$.}
\end{figure*}
In
Ref.~\cite{Butkevich:2020ysf} was shown that this approach describes well the
electron scattering data for carbon, calcium, and argon at different
kinematics. The calculated and measured cross sections are in agreement within
the experimental uncertainties.
 
\section{Results and analysis}

The first measurement of exclusive CCQE-like neutrino-argon interaction cross
sections, performed using the MicroBooNE liquid argon time projector chamber
(LArTPC) neutrino detector was presented in Ref.~\cite{Abratenko:2020sga}.
A specific subset of CCQE-like interactions (CC1p0$\pi$ interactions), includes
CC $\nu_{\mu}-{}^{40}Ar$ scattering events with a detected muon and exactly one
proton, with momenta greater than 100 and 300 MeV/c, respectively. The
data were taken in a phase-space region that corresponds to
$0.1 < p_{\mu} < 1.5$ GeV/c, $0.3 < p_p < 1$ GeV/c,
$-0.65 < \cos\theta < 0.95$, and $\cos\theta_p > 0.15$. After the
application of the event selection requirement only 410 CC1p0$\pi$ candidate
events were retained.

\subsection{Semi-exclusive CCQE double differential cross section}

For these CC1p0$\pi$ events were measured the flux-integrated
$\nu_{\mu}-{}^{40}Ar$ differential cross sections in muon and proton momentum
and angle, and as a function of the calorimetric measured energy and
reconstructed momentum transfer. The statistical uncertainty of the integrated
measured CC1p0$\pi$ cross section is 15.9\% and the systematic uncertainty
sums to 26.2\%. The MicroBooNE detector is located along the
Booster Neutrino Beam at Fermilab. The BNB energy spectrum extends to
2 GeV and peaks around 0.7 GeV~\cite{Aguilar-Arevalo:2008yp}. In this work we
calculate within the RDWIA model with $M_A=1$ GeV and 1.2 GeV the
flux-integrated CCQE semi-exclusive cross sections, taking into account the
MicroBooNE momentum thresholds for muons and protons. Thus, we do not consider
MEC nor the process where charged pions may be produced in the final state.

The flux-integrated double-differential cross section
$d^2\sigma/dp_{\mu}d\cos\theta $ of the semi-exclusive CCQE $\nu_{\mu}-{}^{40}Ar$
scattering is presented in figure~\ref{Fig1}, which shows the cross section
as a function of muon momentum $p_{\mu}$ and muon scattering angle
$\cos\theta$. Here the result was obtained in the RDWIA approach with
the value of the nucleon axial mass $M_A=1$ GeV. The maximum of the calculated
cross section is in the range $0.9 < p_{\mu} < 1.1$ GeV/c
and $0.8 < \cos\theta < 0.96$. So, neutrino interactions with energy
higher than 1 GeV and high values of $\cos\theta$ that corresponds to low
momentum transfer, yield the main contribution to the measured cross sections.

Figures~\ref{Fig2} and~\ref{Fig3} show the flux-integrated
$d^2\sigma/dp_{\mu}d\cos\theta $ cross sections as functions of $p_{\mu}$ for
several bins of the muon scattering angle. On can observe that within the
RDWIA model with $M_A=1.2$ GeV the cross sections in the region of the QE peak
are predicted to be on about 20\% higher than cross sections calculated with
$M_A=1$ GeV.
\begin{figure*}
  \begin{center}
    \includegraphics[height=11cm,width=12cm]{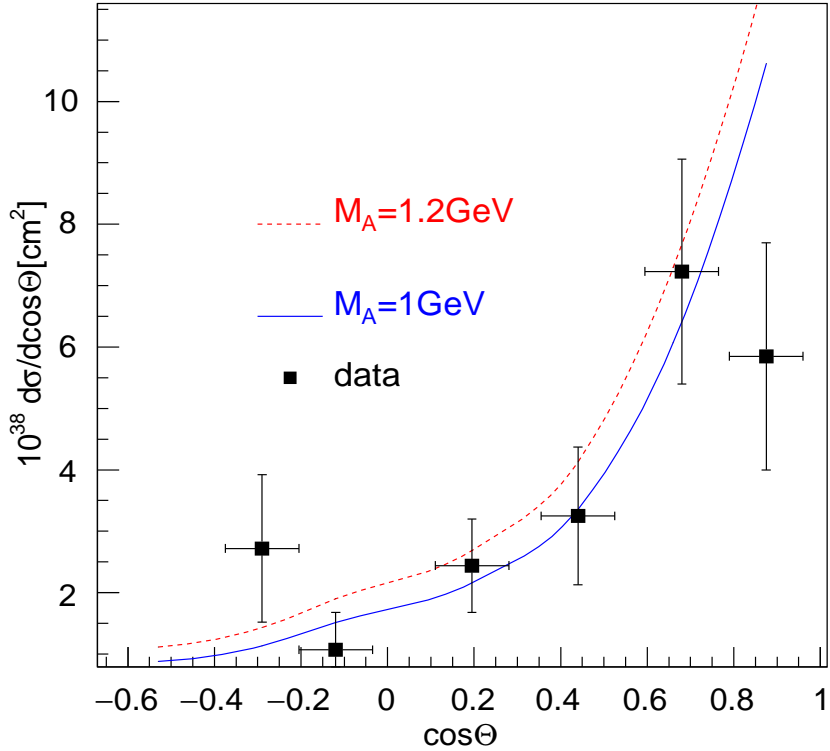}
  \end{center}
  \caption{\label{Fig4} The flux integrated single differential
    $d\sigma/d\cos\theta$ cross section as a function of the cosine of the
    measured muon scattering angle. Error bars show the total (statistical and
    systematic) uncertainty at 68\% confidence level. The colored lines show the
    results of the RDWIA calculation with $M_A=1$ GeV and 1.2 GeV.} 
\end{figure*}
\begin{figure*}
  \begin{center}
    \includegraphics[height=16cm,width=12cm]{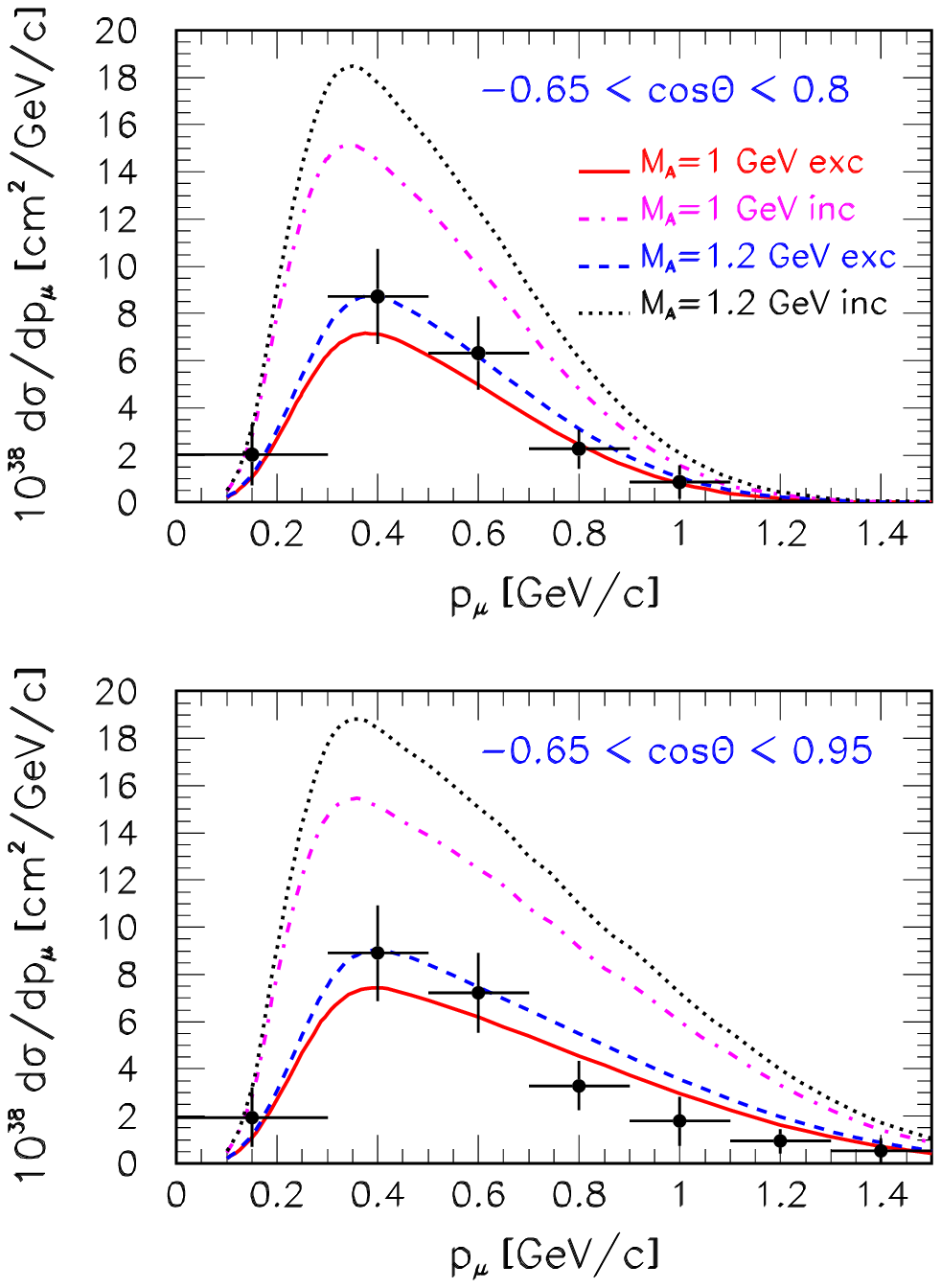}
  \end{center}
  \caption{\label{Fig5}The flux-integrated differential
 $d\sigma/dp_{\mu}$ cross section as a function of muon momentum. Cross sections
    are shown for events with $\cos\theta < 0.8$ (top) and for the full
    measured phase space (bottom). Error bars show the total uncertainty at
    $1 \sigma$ confidence level. As shown in the key the semi-exclusive
    (solid and dashed lines) and inclusive (dash-dotted and dotted lines)
    cross sections were calculated with $M_A=1$ GeV and 1.2 GeV.}
\end{figure*}

\subsection{Semi-exclusive CCQE single differential cross section}

Figure~\ref{Fig4} shows the flux-integrated differential $d\sigma/d\cos\theta$
cross section as a function of the cosine of the measured muon scattering angle.
In Ref.~\cite{Abratenko:2020sga} was shown that the bin migration effects on
this measurement are small and within the assessed uncertainties. The data are
compared to the RDWIA calculations.
\begin{figure*}
  \begin{center}
    \includegraphics[height=16cm,width=12cm]{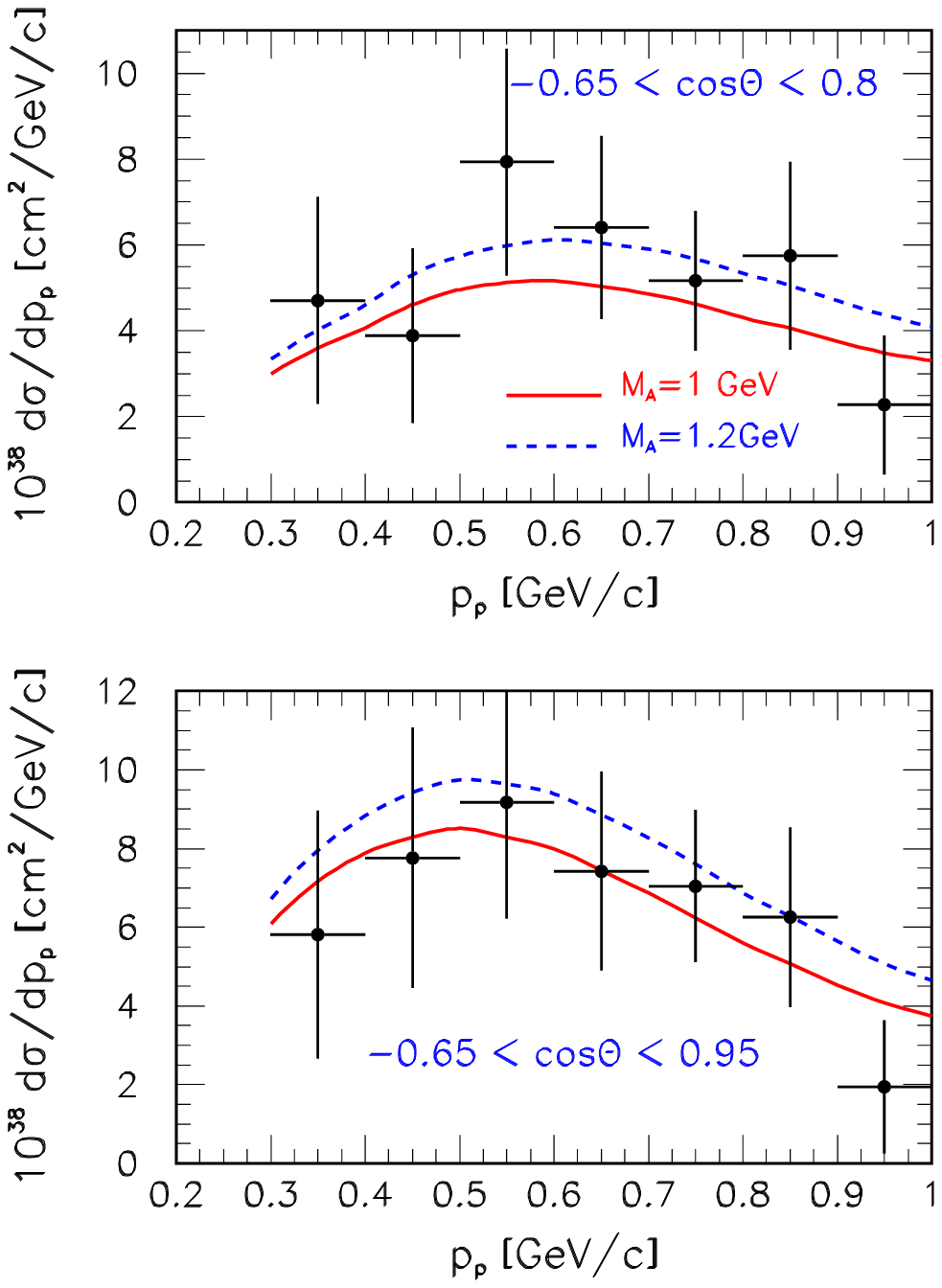}
  \end{center}
  \caption{\label{Fig6}The flux-integrated differential
 $d\sigma/dp_p$ cross section as a function of proton momentum. Cross sections
    are shown for events with $\cos\theta < 0.8$ (top) and for the full
    measured phase space (bottom). Error bars show the total uncertainty at
    $1 \sigma$ confidence level. Colored line show the results of the RDWIA
    calculations with $M_A=1$ GeV and 1.2 GeV.}
\end{figure*}
As can be
seen in figure, calculated cross sections are in overall agreement with data,
except for the highest $\cos\theta$ bin, where the measured cross section is
lower than the theoretical predictions. We note that this bin corresponds to
low-momentum transfer ($Q^2<0.1$ (GeV/c)$^2$), where the nuclear effects
are significant and the neutrino interactions with energy $\var_{\nu} > 1$ GeV
(high energy range of the BNB neutrino flux) give the main contribution to the
measured $d\sigma/d\cos\theta$ cross section.
\begin{figure*}
  \begin{center}
    \includegraphics[height=16cm,width=12cm]{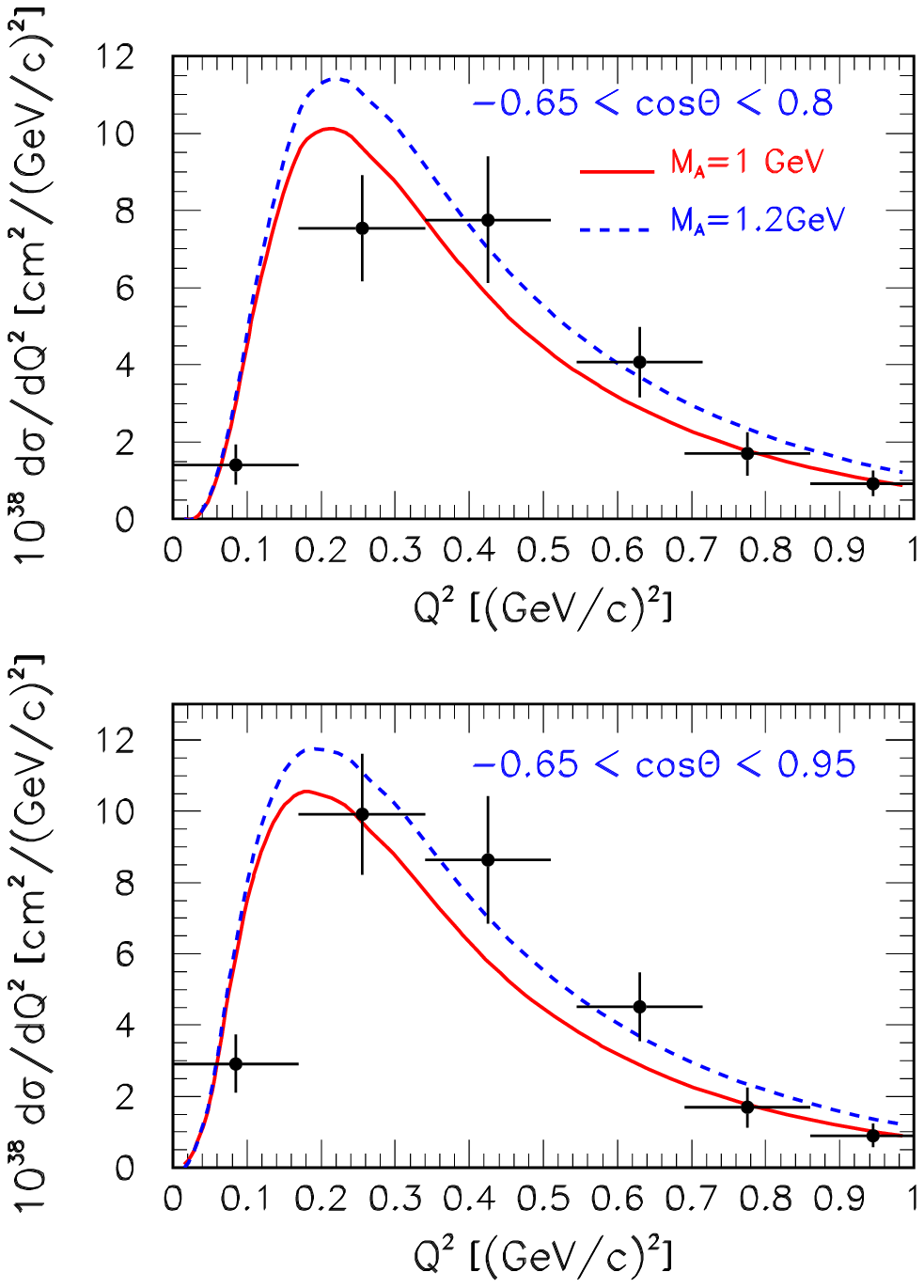}
  \end{center}
  \caption{\label{Fig7} As Figure 6, but for the differential $d\sigma/dQ^2$
  cross section as a function of $Q^2$. }
\end{figure*}

As the differential $d\sigma/dp_{\mu}$, $d\sigma/dp_p$, and $d\sigma/dQ^2$ cross
sections include contributions from all muon scattering angles, their agreement
with the theoretical calculations is effected by these inclusions. In Ref.
\cite{Abratenko:2020sga} the relevant cross sections in the case where events
with $\cos\theta > 0.8$ are excluded and in the full available
phase space $-0.65 < \cos\theta < 0.95$ are presented. Figure~\ref{Fig5} shows
measured differential $d\sigma/dp_{\mu}$ cross section as a function of muon
momentum for $-0.65 < \cos\theta < 0.95$ and $-0.65 < \cos\theta < 0.8$
compared to the RDWIA calculations. Removing events with $\cos\theta > 0.8$
significantly improves the agreement between data and theory at $p_{\mu}>0.6$
GeV/c. The calculated flux-integrated cross sections for inclusive reaction
are shown as well in figure~\ref{Fig5} for the full measured phase
space and for events with $\cos\theta < 0.8$. The contribution of
$(\nu_{\mu}, \mu p)$ channel with $p_p > 300$ MeV/c to the inclusive
$d\sigma/dp_{\mu}$ cross section increases slowly from 35\% at
$p_{\mu}\approx 0.2$ GeV/c to 50\% at $p_{\mu}\approx 1$ GeV/c. The value of muon
momentum where the maximum of $d\sigma/dp_{\mu}$ cross sections appears is
about 0.4 GeV/c.

The differential cross sections $d\sigma/dp_p$ as functions of proton momentum
are shown in Figure~\ref{Fig6} with and without events with
$\cos\theta > 0.8$. Also shown are the results obtained in the RDWIA. Overall,
agreement is observed between data and calculations, even for the full event
sample without the $\cos\theta > 0.8$ requirement. Figure~\ref{Fig6}
demonstrates that the measured proton momentum distribution is wider than the
muon momentum distribution and the maximum in the $d\sigma/dp_{\mu}$ cross
section is located at $p_p=0.5$ GeV/c.

Finally, figure~\ref{Fig7} shows the flux-integrated differential
$d\sigma/dQ^2$ cross sections as functions of $Q^2$ for
$-0.65 < \cos\theta <0.95$ and $-0.65 < \cos\theta <0.8$. The data are
compared to the RDWIA calculations. Note, that in the bin $0.17< Q^2 <0.34$
(GeV/c)$^2$ the agreement between data and theoretical result for the full
phase space is better than for event sample with the $\cos\theta$ requirement.
As can be seen in figure~\ref{Fig7} at $Q^2<0.1$ (GeV/c)$^2$, the measured
cross section is significantly lower than the theoretical prediction. Note,
that at low $Q^2$ the $d\sigma/dQ^2$ cross section depends weakly on the
value of axial mass and $Q^2$ distributions are controled by nuclear effects.

We calculated the $\chi^2$ value for the agreement of the RDWIA prediction
with data as simple sum of those $\chi^2$ values obtained for
$d\sigma/d\cos\theta$, $d\sigma/dp_{\mu}$, and $d\sigma/dQ^2$ distributions
separately.
As follows from this analysis the values of $\chi^2$/degree of
freedom (d.o.f.) for $M_A=1 (1.2)$ GeV are $\chi^2/d.o.f.=1.12$
(1.36) for $\cos\theta < 0.8$ and $\chi^2/d.o.f.=1.26$ (2.22) for
$\cos\theta < 0.95$. The measured integrated cross sections obtained by
integrating $d\sigma/d\cos\theta$ cross section over $-0.64 < \cos\theta < 0.8$
and $-0.64 < \cos\theta < 0.95$ are equal to $(4.05\pm 1.4)\times 10^{-38}$
cm${}^{2}$ and $(4.93\pm 1.55)\times 10^{-38}$ cm${}^{2}$, correspondingly
~\cite{Abratenko:2020sga}. The calculated with $M_A=1 (1.2)$ GeV cross section
values of $3.65\times 10^{-38} (4.48\times 10^{-38})$ cm$^2$ for
$\cos\theta < 0.8$ and $5.24\times 10^{-38} (6.30\times 10^{-38})$ cm$^2$ for
the full measured phase space agree also with data. On the other hand the
statistical and systematic precision of the MicroBooNE data are insufficient for
current needs.
Thus, within the RDWIA approach the measured flux-integrated CC1p0$\pi$
differential and integral cross sections can be described well within the
experimental errors with $1 < M_A < 1.2$ GeV. These values of $M_A$ are in
agreement with the best fit values $M_A=1.15 \pm 0.03$ GeV and
$M_A=1.2 \pm 0.06$ GeV obtained from the CCQE-like fit of the MiniBooNE and
MINERvA data in Refs.~\cite{Wilkinson:2016wmz, Butkevich:2018hll}.
 
\section{Conclusions}
 
In this work we study the semi-exclusive CCQE $\nu_{\mu}{}^{40}$Ar scattering.
Within the RDWIA approach the flux-integrated CC1p0$\pi$ differential and
integral cross sections were calculated with $M_A=1$ GeV and 1.2 GeV.
It was shown that the maximum of the double differential cross sections is in
the range $0.9 < p_{\mu} < 1.1 $ GeV and $0.8 < \cos\theta < 0.95$. The
calculated single differential cross sections were tested against the
MicroBooNE data. We found that the muon angular distribution
is in overall agreement with measured one except at small muon scattering
angle, where the measured cross section is about $2\sigma$ lower than the
theoretical predictions.

The differential cross sections calculated in the RDWIA approach with
$M_A=1$ GeV and 1.2 GeV for $\cos\theta < 0.8$, and with $M_A=1$ GeV for full
phase space are in agreement with data. The calculated integral CC1p0$\pi$
cross sections also agree with data even for the full event sample without
the $\cos\theta < 0.8$ requirement. The contribution of the exclusive
$(\nu_{\mu}, \mu p)$ channel with $p_p>300$ MeV/c to the inclusive cross
sections is about 50\% at $p_p=1$ GeV/c. The measurements of the double and
single differential exclusive CC1p0$\pi$ cross sections on ${}^{40}$Ar with
statistical and systematical uncertainty better than 20\% allow to constrains
models of the CCQE interaction and values of $M_A$ that use in precision
neutrino oscillation analysis.

\section*{Acknowledgments}

The author greatly acknowledges A. Habig for fruitful discussions and a
critical reading of the manuscript.

\bibliographystyle{JHEP}
\bibliography{bibfile}

\providecommand{\href}[2]{#2}\begingroup\raggedright\begin{thebibliography}{100}

\bibitem{Acero:2019ksn}
  {\scshape NOvA} collaboration, M.~A. Acero et~al., \emph{{First Measurement
      of Neutrino Oscillation Parameters using Neutrinos and Antineutrinos by
      NOvA}},
  \href{http://dx.doi.org/10.1103/PhysRevLett.123.151803}{\emph{Phys. Rev.
      Lett.} {\bf 123} (2019) 151803}, [\href{https://arxiv.org/abs/1906.04907}
    {{\tt 1906.04907}}].

 \bibitem{Abe:2020vdv}
   {\scshape T2K} collaboration, K.~Abe et~al., \emph{{T2K measurements of
       muon neutrino and antineutrino disappearance using $3.13\times 10^{21}$
       protons on target}},
     \href{http://dx.doi.org/10.1103/PhysRevD.103.L011101}{\emph{Phys. Rev. D.}
       {\bf 103} (2021) L011101}, [\href{https://arxiv.org/abs/2008.07921}
    {{\tt 2008.07921}}].
 
 \bibitem{Acciarri:2016crz}
   {\scshape DUNE} collaboration, R.~Acciarri et~al., \emph{{Long-Baseline
       Neutrino Facility (LBNF) and Deep Underground Neutrino Experiment
       (DUNE)}: {Conceptual Design Report, Volume 1: The LBNF and DUNE
       Projects}},
  [\href{https://arxiv.org/abs/1601.05471}{{\tt 1601.05471}}].

   \bibitem{Abe:2018uyc}
     {\scshape Hyper-Kamiokande} collaboration, K.~Abe et~al.,
       \emph{{Hyper-Kamiokande Design Report}},
  [\href{https://arxiv.org/abs/1805.04163}{{\tt 1805.04163}}].

 \bibitem{Aguilar-Arevalo:2010zc}
   {\scshape MiniBooNE} collaboration, A.~A.~Aguilar-Arevalo et~al.,
   \emph{{First Measurement of the Muon Neutrino Charged Current Quasielastic
       Double Differential Cross Section}},
     \href{http://dx.doi.org/10.1103/PhysRevD.81.092005}{\emph{Phys. Rev. D.}
       {\bf 81} (2010) 092005}, [\href{https://arxiv.org/abs/1002.2680}
    {{\tt 1002.2680}}].

 \bibitem{Abe:2014iza}
   {\scshape T2K} collaboration, K.~Abe et~al., \emph{{Measurement of the
       \ensuremath{\nu_{\mu}} charged-current quasielastic cross section on
       carbon with the ND280 detector at T2K}},
     \href{http://dx.doi.org/10.1103/PhysRevD.92.112003}{\emph{Phys. Rev. D.}
       {\bf 92} (2015) 112003}, [\href{https://arxiv.org/abs/1411.6264}
    {{\tt 1411.6264}}].

 \bibitem{Fiorentini:2013ezn}
        {\scshape MINERvA} collaboration, G.~A.~Fiorentini et~al.,
        \emph{{Measurement of Muon Neutrino Quasielastic Scattering on a
            Hydrocarbon  Target at $E_\nu \sim 3.5$ GeV}},
        \href{http://dx.doi.org/10.1103/PhysRevLett.111.022502}
             {\emph{Phys. Rev. Lett.}
       {\bf 111} (2013) 022502}, [\href{https://arxiv.org/abs/1305.2243}
    {{\tt 1305.2243}}].

 \bibitem{Walton:2014esl}
        {\scshape MINERvA} collaboration, T.~Walton et~al.,
        \emph{{Measurement of muon plus proton final states in $\nu_{\mu}$
          interactions on hydrocarbon at $< E_{\nu} > = $ 4.2 GeV}},
        \href{http://dx.doi.org/10.1103/PhysRevD.91.071301}
             {\emph{Phys. Rev. D.}
       {\bf 111} (2015) 071301}, [\href{https://arxiv.org/abs/1409.4497}
    {{\tt 1409.4497}}].

 \bibitem{Betancourt:2017uso}
        {\scshape MINERvA} collaboration, M.~Betancourt et~al.,
        \emph{{Direct Measurement of Nuclear Dependence of Charged Current
            Quasielasticlike Neutrino Interactions Using MINER$\nu$A}},
        \href{http://dx.doi.org/10.1103/PhysRevLett.119.082001}
             {\emph{Phys. Rev. Lett.}
       {\bf 119} (2017) 082001}, [\href{https://arxiv.org/abs/1705.03791}
    {{\tt 1705.03791}}].

 \bibitem{Abe:2018pwo}
   {\scshape T2K} collaboration, K.~Abe et~al., \emph{{Characterization of
       nuclear effects in muon-neutrino scattering on hydrocarbon with a
       measurement of final-state kinematics and correlations in
       charged-current pionless interactions}},
     \href{http://dx.doi.org/10.1103/PhysRevD.98.032003}{\emph{Phys. Rev. D.}
       {\bf 98} (2018) 032003}, [\href{https://arxiv.org/abs/1802.05078}
       {{\tt 1802.05078}}].

 \bibitem{Abratenko:2019jqo}
   {\scshape MicroBooNE} collaboration, P.~Abratenko et~al.,
   \emph{{First Measurement of Inclusive Muon Neutrino Charged Current
       Differential Cross Sections on Argon at $E_\nu\sim$0.8 GeV with the
       MicroBooNE Detector}},
   \href{http://dx.doi.org/10.1103/PhysRevLett.123.131801}
        {\emph{Phys. Rev. Lett.}{\bf 123} (2019) 131801},
        [\href{https://arxiv.org/abs/1905.09694}{{\tt 1905.09694}}].     

  \bibitem{Abratenko:2020acr}
   {\scshape MicroBooNE} collaboration, P.~Abratenko et~al.,
   \emph{{First Measurement of Differential Charged Current Quasielastic-like
    $\nu_\mu$-Argon Scattering Cross Sections with the MicroBooNE Detector}},
   \href{http://dx.doi.org/10.1103/PhysRevLett.125.201803}
        {\emph{Phys. Rev. Lett.}{\bf 125} (2020) 201803},
        [\href{https://arxiv.org/abs/2006.00108}{{\tt 2006.00108}}].     

  \bibitem{Abratenko:2020sga}
   {\scshape MicroBooNE} collaboration, P.~Abratenko et~al.,
   \emph{{Measurement of differential cross sections for $\nu_{\mu}$ -Ar
       charged-current interactions with protons and no pions in the final state
       with the MicroBooNE detector}},
   \href{http://dx.doi.org/10.1103/PhysRevD.102.112013}
        {\emph{Phys. Rev. D.}{\bf 102} (2020) 112013},
        [\href{https://arxiv.org/abs/2010.02390}{{\tt 2010.02390}}].     
        
 \bibitem{Antonello:2015lea}
 {\scshape MicroBooNe, LAr1-ND, ICARUS-WA104} collaboration, M.~Antonello
   et~al., \emph{{A Proposal for a Three Detector Short-Baseline Neutrino
   Oscillation Program in the Fermilab Booster Neutrino Beam}},
  [\href{https://arxiv.org/abs/1503.01520}{{\tt 1503.01520}}].

\bibitem{Frullani:1984nn}
  S.~Frullani and J.~Mougey, \emph{{Single Particle Properties of Nuclei
      Through (e, e$^{\prime}$ p) Reactions}},
  {\emph{Adv. Nucl. Phys.} {\bf 14} (1984) 1--283}.

\bibitem{Dieperink:1990uk}
  A.~E.~L.~Dieperink, P.~K.~A.~Huberts, \emph{{On high resolution
      (e,e$^{\prime}$,p) reactions}},
  \href{http://dx.doi.org/10.1146/annurev.ns.40.120190.001323}{\emph{Ann. Rev.
      Nucl. Part. Sci.} {\bf40} (1990) 239--284}.

 \bibitem{Fissum:2004we}
   {\scshape Jefferson Lab Hall A} collaboration, K.~G.~Fissum et~al.,
   \emph{{The Dynamics of the quasielastic ${}^{16}$O(e, e$^{\prime}$ p)
 reaction at $Q^2$ =\textasciitilde{} 0.8 (GeV/c)${}^2$}},
   \href{http://dx.doi.org/10.1103/PhysRevC.70.034606}
   {\emph{Phys. Rev. C.}{\bf 70} (2004) 034606},
   [\href{https://arxiv.org/abs/nucl-ex/0401021}{{\tt nucl-ex/0401021}}].

 \bibitem{Dutta:2003yt}
   {\scshape JLab E91013} collaboration, D.~Dutta et~al.,
   \emph{{A Study of the quasielastic (e,e$^{\prime}$ p) reaction on ${}^{12}$C,
       ${}^{56}$Fe and ${}^{97}$Au}},
   \href{http://dx.doi.org/10.1103/PhysRevC.68.064603}
   {\emph{Phys. Rev. C.}{\bf 68} (2003) 064603},
   [\href{https://arxiv.org/abs/nucl-ex/0303011}{{\tt nucl-ex/0303011}}].

 \bibitem{Gu:2020shc}
   {\scshape Jefferson Lab Hall A} collaboration, L.~Gu et~al.,
   \emph{{Measurement of the Ar(e,e$^\prime$ p) and Ti(e,e$^\prime$ p) cross
   sections in Jefferson Lab Hall A}},
   \href{http://dx.doi.org/10.1103/PhysRevC.103.034604}
   {\emph{Phys. Rev. C.}{\bf 103} (2021) 034604},
   [\href{https://arxiv.org/abs/2012.11466}{{\tt 2012.11466}}].

\bibitem{Nakamura:1974zz}
  K.~Nakamura, S.~Hiramatsu, T~Kamae, H.~Muramatsu, H.~Izutsu, Y.~Watase
  \emph{{Reaction ${}^{40}$Ca(e, e' p) and Observation of the 1s Proton State}},
  \href{http://dx.doi.org/10.1103/PhysRevLett.33.853}{\emph{Phys. Rev. Lett.}
  {\bf 33} (1974) 853--855}.

 \bibitem{Nakamura:1976kn}
  K.~Nakamura, S.~Hiramatsu, T~Kamae, H.~Muramatsu, H.~Izutsu, Y.~Watase,
  \emph{{The ${}^{27}$Al, ${}^{40}$Ca and ${}^{51}$V(e, e$^\prime$ p) Reactions
      and Observation of Deep Hole States}},
  \href{http://dx.doi.org/10.1016/0375-9474(76)90243-8}{\emph{Nucl. Phys. A}
  {\bf 271} (1976) 221--234}.

\bibitem{Mougey:1976sc}
  J.~Mougey, M.~ Bernheim, A.~Bussiere, A.~Gillibert, Xuan~Ho~Phan, M.~Priou,
  D.~Royer, I.~Sick, G.~J.~Wagner,
  \emph{{Quasifree (e, e$^\prime$ p) Scattering on ${}^{12}$C, ${}^{28}$Si,
      ${}^{40}$Ca and ${}^{58}$Ni}},
  \href{http://dx.doi.org/10.1016/0375-9474(76)90510-8}{\emph{Nucl. Phys. A}
  {\bf 262} (1976) 461--492}.

\bibitem{Kramer:1990}
  G.~J. Kramer, \emph{{The proton spectral function of ${}^{40}$Ca and
      ${}^{48}$Ca studing with the (e, e$^\prime$ p) reaction. An investigation
      of ground-state correlations}}.
\newblock PhD thesis, Amsterdam U., 1990 (unpublished).
  
\bibitem{Kramer:1989uiu}
  G.~J.~Kramer et~al.,
  \emph{{Proton ground-state correlations in ${}^{40}$Ca studied with the
      reaction ${}^{}$Ca(e, e$^\prime$ p) ${}^{49}$K}},
  \href{http://dx.doi.org/10.1016/S0370-2693(89)80022-X}{\emph{Phys. Lett. B.}
  {\bf 227} (1989) 199--203}.

\bibitem{Kramer:2000kc}
  G.~J.~Kramer, H.~P.~Blok, Louk~Lapikas,
  \emph{{A Consistent analysis of (e,e$^\prime$ p) and (d,${}^{3}$H)
      experiments}},
  \href{http://dx.doi.org/10.1016/S0375-9474(00)00379-1}{\emph{Nucl. Phys. A}
  {\bf 679} (2001) 267--286},
   [\href{https://arxiv.org/abs/nucl-ex/0007014}{{\tt nucl-ex/0007014}}].

 \bibitem{Kelly:1996hd} J.~J.~Kelly,
   \emph{{Nucleon knockout by intermediate-energy electrons}},
  \href{http://dx.doi.org/10.1007/0-306-47067-5_2}{\emph{Adv. Nucl. Phys.}
  {\bf 23} (1996) 75--294}.
   
\bibitem{Udias:2001tc} J.~M.~Udias and J.~A.~Caballero and
  E.~Moya de Guerra and Javier~R.~Vignote and A.~Escuderos,
  \emph{{Relativistic mean field approximation to the analysis of
      ${}^{16}$0(e, e$^\prime$ p) ${}^{15}$N data at $Q^2$ less than
      0.4(GeV/c)${}^2$}},
  \href{http://dx.doi.org/10.1103/PhysRevC.64.024614}{\emph{Phys. Rev. C}
  {\bf 64} (2001) 024614},
   [\href{https://arxiv.org/abs/nucl-th/0101038}{{\tt nucl-ex/0101038}}].

\bibitem{Kelly:2005is} J.~J.~Kelly,
  \emph{{RDWIA analysis of ${}^{12}$C(e, e$^\prime$ p) for $Q^2$
      \ensuremath{<} 2 (GeV/c)$^2$}},
  \href{http://dx.doi.org/10.1103/PhysRevC.71.064610}{\emph{Phys. Rev. C}
  {\bf 71} (2005) 064610},
   [\href{https://arxiv.org/abs/nucl-th/0501079}{{\tt nucl-ex/0501079}}].
   
 \bibitem{Maieron:2003df}
   C.~Maieron, M.~C.~Martinez, J.~A.~Caballero, J.~M.~Udias,   
   \emph{{Nuclear model effects in charged current neutrino nucleus
       quasielastic scattering}},
  \href{http://dx.doi.org/10.1103/PhysRevC.68.048501}{\emph{Phys. Rev. C}
  {\bf 68} (2003) 048501}.
   [\href{https://arxiv.org/abs/nucl-th/0303075}{{\tt nucl-ex/0303075}}].

 \bibitem{Meucci:2003cv}
   Andrea~Meucci, Carlotta~Giusti, Franco~Davide~Pacati, 
   \emph{{Relativistic Green's function approach to charged current neutrino
       nucleus quasielastic scattering}},
  \href{http://dx.doi.org/10.1016/j.nuclphysa.2004.04.108}{\emph{Nucl. Phys. A}
  {\bf 739} (2004) 277--290},
   [\href{https://arxiv.org/abs/nucl-th/0311081}{{\tt nucl-ex/0311081}}].

 \bibitem{Martinez:2005xe}
   M.~C.~Martinez, P.~Lava, N.~Jachowicz, J.~Ryckebusch, K.~Vantournhout,
   J.~M.~Udias,
   \emph{{Relativistic models for quasi-elastic neutrino scattering}},
  \href{http://dx.doi.org/10.1103/PhysRevC.73.024607}{\emph{Phys. Rev. C}
  {\bf 73} (2006) 024607},
   [\href{https://arxiv.org/abs/nucl-th/0505008}{{\tt nucl-ex/0505008}}].
  
 \bibitem{Picklesimer:1985ey}
   A.~Picklesimer, J.~W.~Van Orden, Stephen~J.~Wallace,
   \emph{{Final State Interactions and Relativistic Effects in the $e$
       (Polarized), $e^\prime p$ Reaction}},
  \href{http://dx.doi.org/10.1103/PhysRevC.32.1312}{\emph{Phys. Rev. C}
  {\bf 32} (1985) 1312--1326}.

   \bibitem{Picklesimer:1986wj}
   A.~Picklesimer, J.~W.~Van Orden,
   \emph{{A Formal Framework for the Electroproduction of Polarized Nucleons
       From Nuclei}},
  \href{http://dx.doi.org/10.1103/PhysRevC.35.266}{\emph{Phys. Rev. C}
  {\bf 35} (1987) 266--279}.

 \bibitem{Kelly:1998ti} J.~J.~Kelly,
   \emph{{Channel coupling in A(polarized e, e$^\prime$ polarized N)B
       reactions}},
  \href{http://dx.doi.org/10.1103/PhysRevC.59.3256}{\emph{Phys. Rev. C}
  {\bf 59} (1999) 3256--3274},
   [\href{https://arxiv.org/abs/nucl-th/9809090}{{\tt nucl-ex/9809090}}].

 \bibitem{Gonzalez-Jimenez:2021ohu}
   R.~Gonzalez-Jimenez, M.~B.~ Barbaro, J.~A.~Caballero, T.~W.~Donnelly,
   N.~Jachowicz, G.~D.~Megias, K.~Niewczas, A.~Nikolakopoulos,
   J.~W.~Van Orden, J.~M.~Udias,
   \emph{{Neutrino energy reconstruction from semi-inclusive samples}}
  [\href{https://arxiv.org/abs/2104.01701}{{\tt 2104.01701}}].

 \bibitem{Amaro:2021sec}
   J.~E.~Amaro, M.~B.~Barbaro, J.~A.~Caballero, T.~W.~Donnelly,
   R.~Gonzalez-Jimenez, G.~D.~Megias, and I.~Ruiz~Simo
   \emph{{Neutrino-Nucleus scattering in the SuSA model}}
  [\href{https://arxiv.org/abs/2106.02857}{{\tt 2106.02857}}].

   \bibitem{Aguilar-Arevalo:2008yp}
   {\scshape MiniBooNE} collaboration, A.~A.~Aguilar-Arevalo et~al.,
   \emph{{The Neutrino Flux prediction at MiniBooNE}},
     \href{http://dx.doi.org/10.1103/PhysRevD.79.072002}{\emph{Phys. Rev. D.}
       {\bf 79} (2009) 072002}, [\href{https://arxiv.org/abs/0806.1449}
    {{\tt 0806.1449}}].

 \bibitem{Butkevich:2007gm}
   A.~V.~Butkevich, Sergey.~A.~Kulagin,
   \emph{{Quasi-elastic neutrino charged-current scattering cross sections on
       oxygen}},
  \href{http://dx.doi.org/10.1103/PhysRevC.76.045502}{\emph{Phys. Rev. C}
  {\bf 76} (2007) 045502},
   [\href{https://arxiv.org/abs/0705.1051}{{\tt 0705.1051}}].

 \bibitem{Butkevich:2009cp}
   A.~V.~Butkevich,
   \emph{{Quasi-elastic neutrino charged-current scattering off C-12}},
  \href{http://dx.doi.org/10.1103/PhysRevC.80.014610}{\emph{Phys. Rev. C}
  {\bf 80} (2009) 014610},
   [\href{https://arxiv.org/abs/0904.1472}{{\tt 0904.1472}}].

 \bibitem{Butkevich:2010cr}
   A.~V.~Butkevich,
   \emph{{Analysis of flux-integrated cross sections for quasi-elastic
       neutrino charged-current scattering off $^{12}$C at MiniBooNE energies}},
  \href{http://dx.doi.org/10.1103/PhysRevC.82.055501}{\emph{Phys. Rev. C}
  {\bf 82} (2010) 055501},
   [\href{https://arxiv.org/abs/1006.1595}{{\tt 1006.1595}}].

 \bibitem{Butkevich:2012zr}
   A.~V.~Butkevich,
   \emph{{Quasi-elastic neutrino charged-current scattering off medium-heavy
       nuclei: ${}^{40}$Ca and ${}^{40}$Ar}},
  \href{http://dx.doi.org/10.1103/PhysRevC.85.065501}{\emph{Phys. Rev. C}
  {\bf 85} (2012) 065501},
   [\href{https://arxiv.org/abs/1204.3160}{{\tt 1204.3160}}].

  \bibitem{Butkevich:2020ysf}
   A.~V.~Butkevich, S.~V.~Luchuk,
   \emph{{Inclusive electron scattering off $^{12}$C, $^{40}$Ca , and $^{40}$Ar :
       Effects of the meson exchange currents}},
  \href{http://dx.doi.org/10.1103/PhysRevC.102.024602}{\emph{Phys. Rev. C}
  {\bf 102} (2020) 024602},
   [\href{https://arxiv.org/abs/2004.04780}{{\tt 2004.04780}}].
  
\bibitem{DeForest:1983ahx}
  T. De Forest, \emph{{Off-Shell electron Nucleon Cross-Sections. The Impulse
      Approximation}},
  \href{http://dx.doi.org/10.1016/0375-9474(83)90124-0}{\emph{Nucl. Phys. A.}
    {\bf 392} (1983) 232--248}.

  \bibitem{Mergell:1995bf}
    P.~Mergell, Ulf G.~Meissner, D.~Drechsel,
    \emph{{Dispersion theoretical analysis of the nucleon electromagnetic
        form-factors}},
  \href{http://dx.doi.org/10.1016/0375-9474(95)00339-8}{\emph{Nucl. Phys. A.}
    {\bf 596} (1996) 367--396},
   [\href{https://arxiv.org/abs/hep-ph/9506375}{{\tt hep-ph/9506375}}].

 \bibitem{Horowitz:1981xw}
 C.~J.~Horowitz, Brian~D.~Serot,
 \emph{{Selfconsistent Hartree Description of Finite Nuclei in a Relativistic
     Quantum Field Theory}},
  \href{http://dx.doi.org/10.1016/0375-9474(81)90770-3}{\emph{Nucl. Phys. A.}
    {\bf 368} (1981) 503--528}.

   \bibitem{Horowitz:1991}
 C.~J.~Horowitz, D.~P.~Murdock, B~D.~Serot,
 \emph{{The Relativistic Impulse Approximation}}, In: K.~Langanke,
 J.~A.~Maruhn, S.~E.~Koonin (eds) {\emph{Computation Nuclear. Physics. 1: }},
 Springer, Berlin, Heidelberg (1991), pg. 129,
 \href{https://doi.org/10.1007/978-3-642-76356-4_7}.

 \bibitem{Cooper:1993nx}
   E.~D.~Cooper, S.~Hama, B.~C.~Clark, R.~L.~Mercer, 
   \emph{{Global Dirac phenomenology for proton nucleus elastic scattering}},
  \href{http://dx.doi.org/10.1103/PhysRevC.47.297}{\emph{Phys. Rev. C}
  {\bf 47} (1993) 297--311}.
   
 
   \bibitem{Kelly:1996} J.~J.~Kelly,
  [\href{http://www.physics.umd.edu/enp/jjkelly/LEA}
    {{\tt http://www.physics.umd.edu/enp/jjkelly/LEA}}].         

 \bibitem{CiofidegliAtti:1995qe}
   C.~Ciofi degli Atti, S.~Simula,
   \emph{{Realistic model of the nucleon spectral function in few and many
    nucleon systems}},
  \href{http://dx.doi.org/10.1103/PhysRevC.53.1689}{\emph{Phys. Rev. C}
  {\bf 53} (1996) 1689}.
   [\href{https://arxiv.org/abs/nucl-th/9507024}{{\tt nucl-th/9507024}}].

 \bibitem{Wilkinson:2016wmz}
   C.~Wilkinson et~al.,
   \emph{{Testing charged current quasi-elastic and multinucleon interaction
       models in the NEUT neutrino interaction generator with published datasets
       from the MiniBooNE and MINER\ensuremath{\nu}A experiments}},
     \href{http://dx.doi.org/10.1103/PhysRevD.93.072010}{\emph{Phys. Rev. D.}
       {\bf 93} (2016) 072010}, [\href{https://arxiv.org/abs/1601.05592}
    {{\tt 1601.05592}}].

 \bibitem{Butkevich:2018hll}
   A.~V.~Butkevich, and S.~V.~Luchuk,
   \emph{{Testing of quasi-elastic neutrino charged-current and two-body meson
       exchange current models with the MiniBooNE neutrino data and analysis
       of these processes at energies available at the NOvA experiment}}
     \href{http://dx.doi.org/10.1103/PhysRevD.99.093001}{\emph{Phys. Rev. D.}
       {\bf 99} (2019) 093001}, [\href{https://arxiv.org/abs/1812.11073}
    {{\tt 1812.11073}}].

\end{thebibliography}\endgroup

\end{document}